\documentclass[floatfix,aps,prd,preprintnumbers,nofootinbib,superscriptaddress]{revtex4}

\pdfoutput=1
\usepackage{amsmath}
\usepackage{amssymb}
\usepackage{slashed}
\usepackage{mathtools}
\usepackage[utf8]{inputenc}
\usepackage{color}
\usepackage[usenames,dvipsnames]{xcolor}
\usepackage[normalem]{ulem}
\usepackage{soul}
\usepackage{units}
\usepackage{rotating}
\usepackage{hhline,multirow,tabularx}
\usepackage{subfigure}
\usepackage{commath}
\usepackage{graphicx,bm}
\usepackage{verbatim}

\newcommand{\T}[1]{\boldsymbol{#1}_{\text{T}}}

\newcommand\3[1]{\boldsymbol{#1}}
\newcommand{\Tsc}[2]{#1_{#2\text{T}}}
\newcommand{\Tscsq}[2]{#1^2_{#2\text{T}}}

\newcommand{\no}{\nonumber \\}
\newcommand{\parz}[1]{\ensuremath{\left(#1\right)}}
\newcommand{\order}[1]{\ensuremath{O\parz{#1}}}

\newcommand{\diff}[1]{\mathrm{d}#1}

\newcommand{\eref}[1]{Eq.~(\ref{e.#1})}
\newcommand{\erefs}[2]{Eqs.~(\ref{e.#1})--(\ref{e.#2})}
\newcommand{\fref}[1]{Fig.~\ref{f.#1}}

\newcommand{\aref}[1]{Appendix~\ref{a.#1}}
\newcommand{\sref}[1]{Sec.~\ref{s.#1}}
\newcommand{\ssref}[1]{Section~\ref{ss.#1}}

\begin{document}

\title{Collinear Factorization in Wide-Angle Hadron Pair Production in 
$e^+e^-$ Annihilation}

\preprint{JLAB-THY-19-3020}
\author{E.~Moffat}
\email{emoff003@odu.edu}
\affiliation{Department of Physics, Old Dominion University, Norfolk, VA 23529, USA}
\author{T.~C.~Rogers}
\thanks{Electronic address: trogers@odu.edu - \href{https://orcid.org/0000-0002-0762-0275}{ORCID: 0000-0002-0762-0275}} 
\affiliation{Department of Physics, Old Dominion University, Norfolk, VA 23529, USA}
\affiliation{Jefferson Lab, 12000 Jefferson Avenue, Newport News, VA 23606, USA}
\author{N.~Sato}
\email{nsato@jlab.org}
\affiliation{Department of Physics, Old Dominion University, Norfolk, VA 23529, USA}
\affiliation{Jefferson Lab, 12000 Jefferson Avenue, Newport News, VA 23606, USA}
\author{A.~Signori}
\thanks{Electronic address: asignori@anl.gov - \href{https://orcid.org/0000-0001-6640-9659}{ORCID: 0000-0001-6640-9659}} 
\affiliation{Physics Division, Argonne National Laboratory, 9700 S. Cass Avenue, Lemont, IL 60439 USA}
\date{August 23, 2019}

%%%%%%%%%%%%%%%%%%%%%%%%%%%%%%%%%%%%%%%%%%%%%%%%%%%
\begin{abstract}
We compute the inclusive unpolarized dihadron production cross section
in the far from back-to-back region of $e^+ e^-$ annihilation in
leading order pQCD using existing fragmentation function fits and standard collinear factorization, focusing on the large transverse momentum region where transverse momentum is comparable to the hard scale (the center-of-mass energy). We compare with standard transverse-momentum-dependent (TMD) fragmentation function-based predictions intended for the small transverse momentum region with the aim of testing the expectation that the two types of calculation roughly coincide at intermediate transverse momentum. 
We find significant tension, within the intermediate transverse momentum region, between calculations done with existing non-perturbative TMD fragmentation functions and collinear factorization calculations if the center-of-mass energy is not extremely large.  
We argue that $e^+ e^-$ measurements are ideal 
for resolving this tension and exploring the large-to-small transverse momentum transition, given the typically larger hard scales ($\gtrsim 10$~GeV)
of the process as compared with similar scenarios that arise in semi-inclusive deep inelastic scattering and fixed-target Drell-Yan measurements. 
\end{abstract}
%%%%%%%%%%%%%%%%%%%%%%%%%%%%%%%%%%%%%%%%%%%%%%%%%%%

\maketitle

%%%%%%%%%%%%%%%%%%%%%%%%%%%%%%%%%%%%%%%%%%%%%%%%%%%
\section{Introduction}
\label{s.intro}

The annihilation of lepton pairs into hadrons is one of a class
of processes notable for being especially clean electromagnetic
probes of elementary quark and gluon correlation functions like parton
density and fragmentation functions (pdfs and ffs)~\cite{Collins:1981uw}. Other such
processes include  inclusive and semi-inclusive deep inelastic
scattering (DIS and SIDIS), and the Drell-Yan (DY) process. In
combination they provide some of the strongest tests of QCD
factorization. However, the exact type of correlation functions involved (e.g.,
transverse momentum dependent, collinear, etc) depends on the
details of the process under consideration and the particular
kinematical regime being accessed.  It is important to confirm the
applicability of each expected factorization for each region, not only
at the largest accessible energies, but also in more moderate energy
regimes, since the latter are especially useful for probing the
non-perturbative details of partonic correlation functions like pdfs
and ffs, and for probing the intrinsic partonic structure of hadrons
generally~\cite{Dudek:2012vr,Accardi:2012qut}. 

In the case of the inclusive lepton-antilepton annihilation into a
dihadron pair, the type of partonic correlation functions accessed
depends on the pair's specific kinematical configuration. In the
back-to-back configuration, there is sensitivity to the intrinsic
non-perturbative transverse momentum of each observed
hadron relative to its parent parton.  This is the regime of
transverse momentum dependent (TMD) factorization, in which TMD ffs
are the relevant correlation functions~\cite{Collins:1981uw,
Collins:1981uk, Collins:1984kg, Collins:2011qcdbook,
GarciaEchevarria:2011rb}.  The TMD region has attracted especially strong interest in phenomenological work in recent decades for its potential to probe
the intrinsic non-perturbative motion of
partons~\cite{Qiu:2000hf,Anselmino:2007fs, Anselmino:2008jk, Anselmino:2013vqa, Signori:2013mda, Anselmino:2013lza, Angeles-Martinez:2015sea, Kang:2015msa, Anselmino:2015fty, Anselmino:2015sxa, Bacchetta:2015ora, Bacchetta:2017gcc, Bertone:2019nxa,Vladimirov:2019bfa} and, more recently, its potential to impact also high-energy measurements~\cite{Nadolsky:2004vt,Angeles-Martinez:2015sea,Bacchetta:2018lna,Bozzi:2019vnl,Martinez:2019mwt,Lupton:2019mwd}. 
See also Refs.~\cite{Boer:2008fr,Metz:2016swz,Vossen:2018nkg} for additional discussions of motivations to study $e^+ e^-$ annihilation into back-to-back hadrons generally, and especially including studies of spin and polarization effects.
If instead the hadrons are nearly collinear, they
can be thought of as resulting from a single hadronizing parent
parton. In that case, the correct formalism uses dihadron
ffs~\cite{Collins:1993kq, Jaffe:1997hf, Courtoy:2012ry,Matevosyan:2018icf}, which are
useful for extracting the transversity pdf without the need for TMD
factorization~\cite{Radici:2001na,Radici:2015mwa,Radici:2018iag}. Finally, if the hadrons are
neither aligned, nor back-to-back, but instead have a large invariant
mass, then the relevant factorization is standard collinear
factorization with collinear ffs~\cite{Sato:2019yez,Ethier:2017zbq,Borsa:2017vwy,Bertone:2017tyb,Bertone:2018ecm,deFlorian:2014xna}   
which has played a significant role in recent years to explore flavor separation in  collinear pdfs using SIDIS data~\cite{Sato:2019yez,Ethier:2017zbq,Borsa:2017vwy}.

Having a fully complete picture of partonic correlation functions and the
roles they play in transversely differential cross sections generally
requires an understanding of the boundaries
between the kinematical regions where different types of factorization
apply and the extent to which those regions overlap~\cite{Arnold:1990yk,Berger:2004cc,Collins:2016hqq,Echevarria:2018qyi}. In this paper, we
focus on the last of the lepton-antilepton annihilation regions
mentioned in the previous paragraph, wherein pure collinear
factorization is expected to be adequate for describing the large deviations from the back-to-back orientation
of the hadron pair. We view this as a natural
starting point for mapping out the regions of the process generally,
since it involves only well-established collinear factorization
theorems and starts with tree-level perturbation theory calculations. It is also motivated by tension between measurements and collinear factorization that has 
already been seen 
in transversely differential 
SIDIS~\cite{Daleo:2004pn,Kniehl:2004hf,Boglione:2014oea,Boglione:2014qha,Su:2014wpa,Gonzalez-Hernandez:2018ipj,Wang:2019bvb}  and 
DY~\cite{Bacchetta:2019tcu}.  That all these cases involve $Q \lesssim 14$~GeV hints that the 
origin of the tension lies with the smaller hard scales. The lack of smooth transition in the intermediate transverse momentum 
region suggests a more complicated than expected role for non-perturbative transverse momentum in the description 
of the large transverse momentum tail when $Q$ is not extremely large. We will elaborate on these issues further in the main text and 
comment on potential resolutions in the conclusion. 

Of course, much work has been done calculating distributions for
this and similar processes, especially in the construction and
development of Monte Carlo event generators~\cite{Sjostrand:2006za,Sjostrand:2014zea,Alwall:2014hca,Bahr:2008pv,Bellm:2015jjp,Bothmann:2019yzt,Schnell:2015gaa,Hautmann:2014kza}.
Our specific interest, however, is in the extent to which the most
direct applications of QCD factorization theorems, with ffs extracted
from other processes, give reasonable behavior 
in the far from
back-to-back region. 
Despite the simplicity of the leading order 
(LO) cross section, it has not, to our knowledge, 
been explicitly presented elsewhere or used in a detailed examination of the transverse momentum dependence of inclusive hadron pairs at wide angle in ordinary collinear pQCD calculations and using standard fragmentation functions.
One challenge to performing such a study is a
dearth of unpolarized dihadron data with transverse momentum dependence for the exact process
under consideration here. In the absence of data, an alternative way to assess the reasonableness of large transverse momentum calculations, and to estimate the point of transition to 
small transverse momentum, is to examine how accurately they match to 
small or medium transverse momentum calculations performed using TMD-based methods, for which many phenomenological results already exist (see e.g. Refs.~\cite{Anselmino2016,Bacchetta:2016ccz,Aschenauer:2015ndk,Boglione:2015zyc,Diehl:2015uka,Rogers:2015sqa,Garzia:2016kqk,Avakian:2016rst} and references therein).

We follow this latter approach in the present paper.  
Namely, using the lowest order (LO) calculation of the far from back-to-back cross section along with standard ff fits~\cite{deFlorian:2014xna}, and comparing with Gaussian-based (or similar) fits 
from, for example, Ref.~\cite{Bacchetta:2017gcc}, we are able to confirm that the two methods of calculation approach one another at intermediate transverse momentum in the very large $Q$ limit, albeit rather slowly. 
At both smaller and larger $Q$, the comparison between 
TMD and collinear based calculations suggests a transition point of between about $\Tsc{q}{}/q_{\rm T}^{\rm Max} \approx .3$ and $.2$, where $q_{\rm T}^{\rm Max}$ is the kinematical maximum of transverse momentum. 
However, at moderate $Q$ of around 12 GeV, the shape of the TMD-based calculation deviates significantly from the collinear at intermediate transverse momentum, and numerically the disagreement at intermediate transverse momentum rises to a factor of several in most places, with the fixed order collinear calculation undershooting the TMD-based calculation. 
This is noteworthy given the similar mismatch with actual data that has been seen in Drell-Yan and SIDIS, already remarked upon above. Whether the solution to the difficulties at moderate transverse momentum lies with the 
collinear treatment or with the phenomenology of TMD functions remains to be seen. 
But all of these observations, we argue, provide enhanced motivation for experimental studies of dihadron pair production that probe the intermediate transition region of the transverse momentum dependence.

We have validated our very large $Q$ and moderate transverse momentum calculation by comparing with transverse momentum distributions 
generated with the default settings of PYTHIA
8~\cite{Sjostrand:2006za,Sjostrand:2014zea}. We find reasonable agreement 
with the PYTHIA generated distributions when the center-of-mass 
energy $Q$ is large ($\sim 50$~GeV). This is perhaps not surprising given that fits of collinear fragmentation 
functions are also generally constrained by large $Q$ measurements. Nevertheless, the specificity of the process makes it a non-trivial consistency validation. 
At lower $Q$ ($\lesssim 10$~GeV) there is much larger disagreement with the event 
generator data, and we comment briefly on the interpretation of this
in the text.

The organization of sections is as follows. In \sref{setup} we set up the basic kinematical description of electron-positron annihilation to two hadrons.  In \ssref{FO} we explain the steps of the LO collinear
calculation at large transverse momentum, in \ssref{asy} we discuss its asymptotically small transverse momentum behavior, and in \ssref{smallqt} we review the basics of the (non-)perturbative TMD calculation for small transverse momentum. We elaborate on our expectations for the validity of the collinear factorization calculation in 
\sref{qthardness}, and in \sref{analysis} we compare and contrast the results at moderate transverse momentum. We comment on these observations and discuss their implications in \sref{conclusion}.

%%%%%%%%%%%%%%%%%%%%%%%%%%%%%%%%%%%%%%%%%%%%%%%%%%%
\section{Kinematical Setup}
\label{s.setup}

The specific process that is the central topic of this paper is semi-inclusive lepton-antilepton (usually electron-
positron) annihilation (SIA) with two observed final-state hadrons:
\begin{equation}
e^-(l)+e^+(l')\rightarrow{H_A}(p_A)+H_B(p_B)+X \, , \label{e.theprocess}
\end{equation}
with a sum over all other final state particles $X$. The $p_A$ and $p_B$ label the momenta of the observed final state 
hadrons, and throughout this paper we will neglect their masses, since we assume hadron masses are negligible relative to hard scales under consideration here.
Our aim is to calculate the cross 
section for this process, differential in the relative transverse momentum of the final state hadron pair, 
and for this  there are a number of useful reference frames. We will mainly follow 
the conventions in Ref.~\cite[13.1-13.2]{Collins:2011qcdbook}.
As indicated in \eref{theprocess}, $l$ and $l'$ will 
label the incoming lepton and antilepton momenta. These 
annihilate to create a highly virtual timelike photon 
with momentum labeled $q$. It is $$Q^2 \equiv q^2$$ that 
sets the hard scale of the process. See also Refs.~\cite{Boer:2008fr,Boer:1997mf} for details on the 
kinematical setup of $e^+ e^-$-annihilation. 
Two particularly useful reference frames are discussed in the next two paragraphs. 

%-------------------------------------------------%
\subsection{Photon frame}

A photon frame is a center-of-mass frame wherein the momenta, in Minkowski coordinates and neglecting masses, are:
\begin{subequations}
\begin{align}
q_\gamma^\mu &= (Q,\3{0}) \, , \\
p_{A,\gamma}^\mu &= \left|\3{p}_{A,\gamma}\right|(1,\3{n}_{A,\gamma}) \, ,  \\
p_{B,\gamma}^\mu &= \left|\3{p}_{B,\gamma}\right|(1,\3{n}_{B,\gamma}) \, . 
\end{align}
\end{subequations}
Here $\3{n}_{A,\gamma}$ and $\3{n}_{B,\gamma}$ are unit vectors in the
directions of the hadron momenta. 
We also define the following unit four-vectors~\cite{Collins:2011qcdbook}:
\begin{equation}
\label{e.unitvects}
Z_\gamma^\mu 
= \frac{(0,\3{n}_{A,\gamma}-\3{n}_{B,\gamma})}{|\3{n}_{A,\gamma}-\3{n}_{B,\gamma}|} \, , \ \ \ \ \ 
X_\gamma^\mu = \frac{(0,\3{n}_{A,\gamma}+\3{n}_{B,\gamma})}{|\3{n}_{A,\gamma}+\3{n}_{B,
\gamma}|} \, .
\end{equation}
The $z$-axis can be fixed to align
along the spatial components of $Z_\gamma^\mu$ and the $x$-axis along the spatial components of $X_\gamma^\mu$. The $z$-axis then
bisects the angle (called $\delta\theta$ in the figure) between
$\3{p}_{A,\gamma}$ and $-\3{p}_{B,\gamma}$.  See
\fref{frames} (A) for an illustration. This is analogous 
to the Collins-Soper frame~\cite{Collins:1977iv} frequently used in Drell-Yan
scattering, where the lepton pair is in the final state.
Another sometimes useful photon rest frame is one in which the spatial $z$-axis lies along the direction of one of the hadrons. This is the 
analogue of the Gottfried-Jackson frame~\cite{Gottfried:1964nx}.
\begin{figure}
\centering
\includegraphics[trim= 0in 0.75in 0in 0.75in,clip,width=0.48\textwidth]
{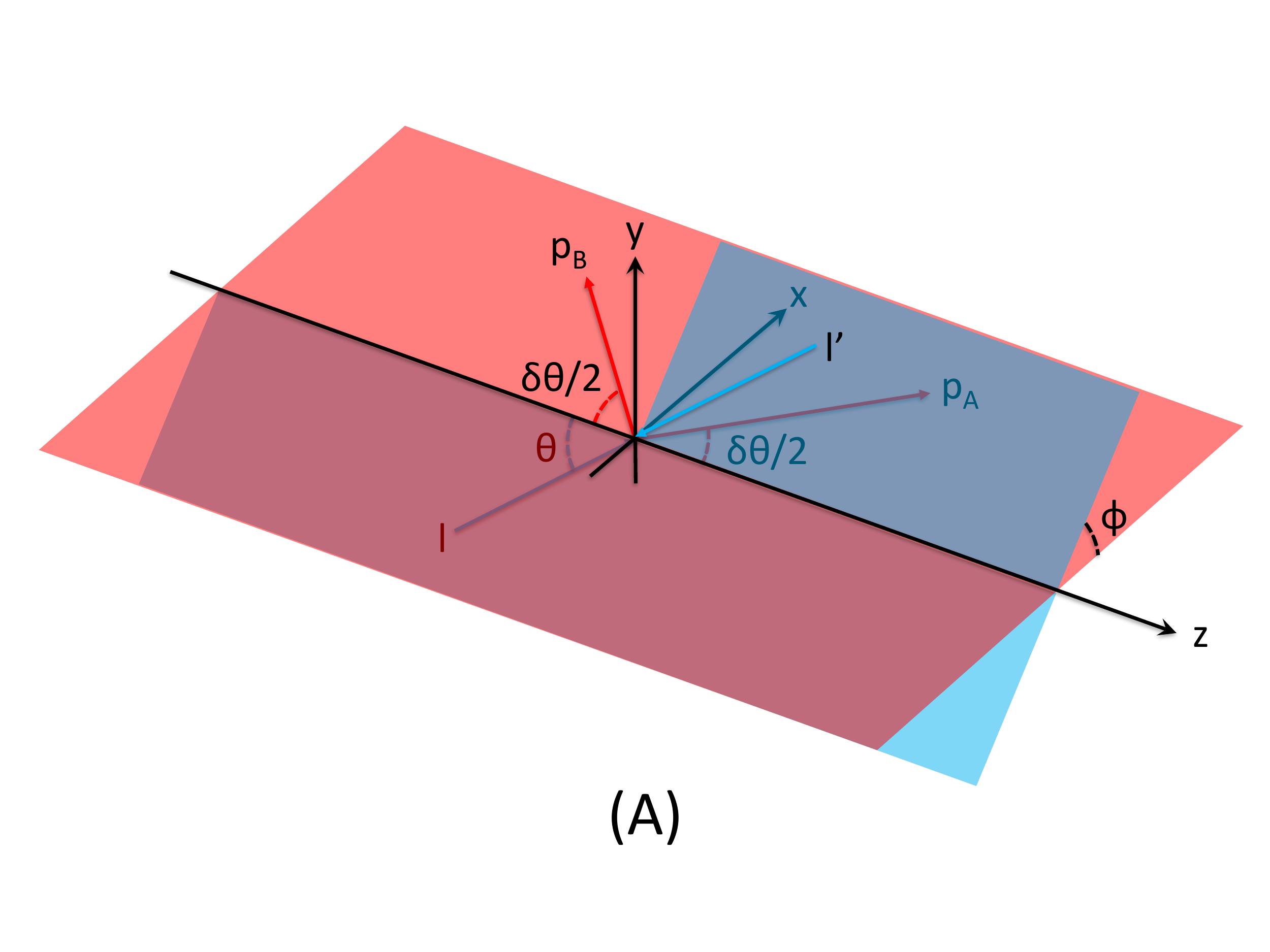}
\includegraphics[trim= 0in 0.75in 0in 0.75in,clip,width=0.48\textwidth]
{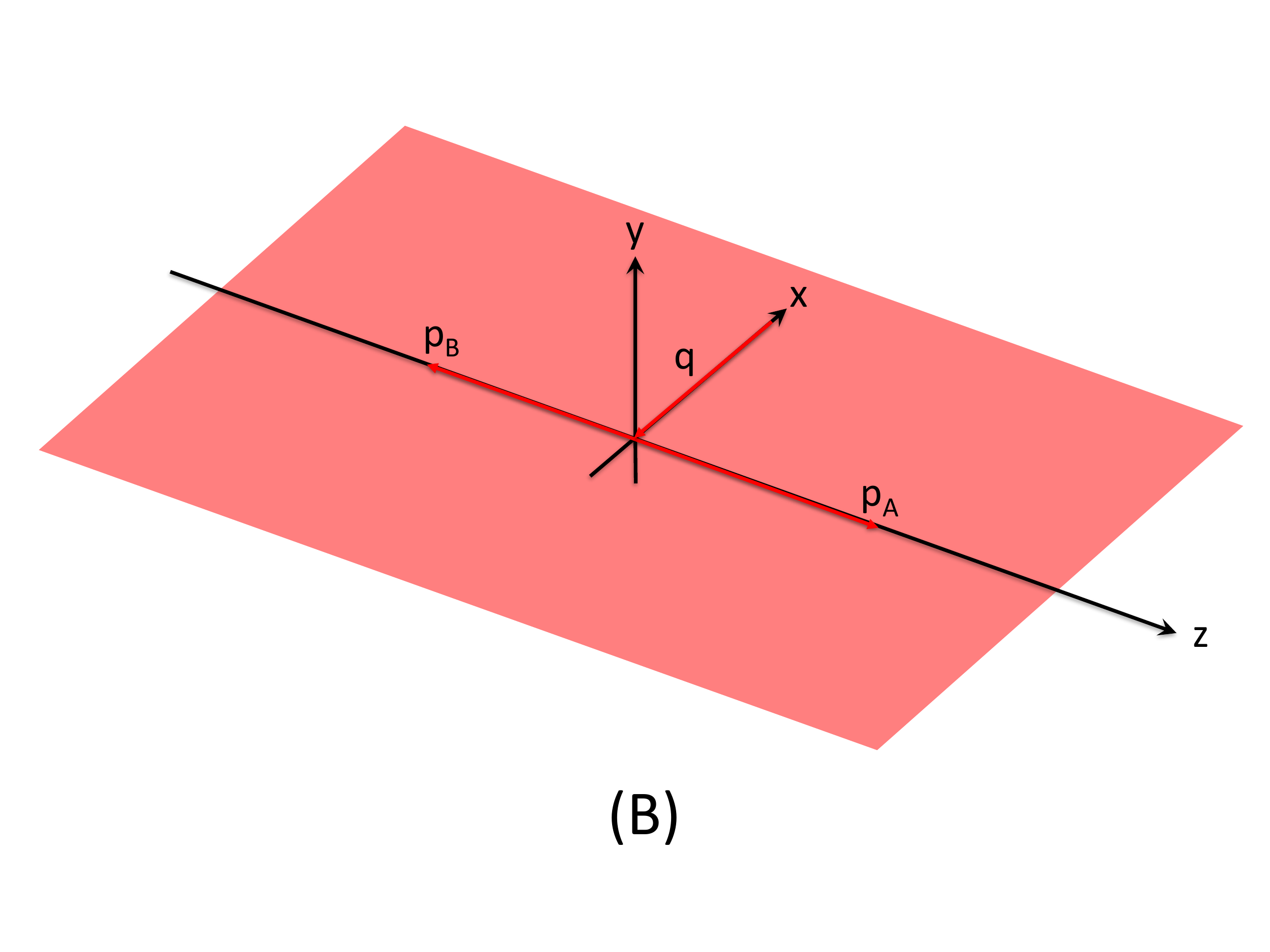}
\caption{(A) The photon frame. The $x$ and $z$ axes have been aligned
  with the spatial components of $X^\mu$ and $Z^\mu$ from
  \eref{unitvects}. The blue plane is the $e^+ e^-$ plane.  (B) The
  hadron frame, with the hadrons exactly back-to-back. See text for further explanation.}
\label{f.frames}
\end{figure}

%-------------------------------------------------%
\subsection{Hadron frame}

In the hadron frame, $p_A$ and $p_B$ are back-to-back along the $z$
axis -- see \fref{frames} (B). The measure of the 
deviation from the back-to-back configuration is then the
size of the virtual photon's transverse momentum, 
$\Tsc{q}{h}$. In light-cone coordinates and neglecting masses the momenta in 
the hadron frame are: 
\begin{subequations}
\begin{align}
q_h &= \left(\sqrt{\frac{Q^2+\Tscsq{q}{h}}{2}},\sqrt{\frac{Q^2+\Tscsq{q}{h}}{2}},\Tsc{\3q}{h}\right)\, , \\  
p_{A,h} &= (p_{A,h}^+,0,\3{0})\, , \\ 
p_{B,h} &= (0,p_{B,h}^-,\3{0}) \, .
\end{align}
\end{subequations}
We have chosen to boost along the $z$-axis in the hadron 
frame until $q_h^+$ = $q_h^-$.
Useful Lorentz-invariant variables are 
\begin{equation}
z_A=\frac{p_A\cdot{p}_B}{q\cdot{p}_B} = \frac{p_{A,h}^+}{q_h^+} \, , \qquad 
z_B=\frac{p_A\cdot{p}_B}{q\cdot{p}_A} = \frac{p_{B,h}^-}{q_h^-}  \, . 
\label{e.zdefs}
\end{equation}
Note that we take the Lorentz invariant ratios
to define $z_A$ and $z_B$. Since in this paper we 
assume that the hadron masses are negligible, these are also equal 
to the light-cone ratios shown. For a treatment 
that includes kinematical mass effects, see Ref.~\cite{Mulders:2019mqo}. 
The transverse momentum of the photon in the hadron frame is: 
\begin{equation}
\Tscsq{q}{h}=\frac{2\ p_A\cdot{q}\ p_B\cdot{q}}{p_A\cdot{p}_B}-Q^2 = Q^2 \tan^2 \parz{\delta \theta/2} \, .
\label{e.qTdef}
\end{equation}
As $\delta \theta$ approaches 180$^{\circ}$ in
\fref{frames}, far from the back-to-back configuration, 
$\Tsc{q}{h}$ as defined in \eref{qTdef} diverges, while 
for $\delta \theta \approx 0$ it approaches zero. From here 
forward, we will drop the $h$ subscript for simplicity 
and $\Tsc{q}{}$ will be understood to refer to the hadron
frame photon transverse momentum. 

The transverse momentum has an absolute kinematical upper bound: 
\begin{equation}
{q_{\rm T}^{\rm Max}}^2 \leq \frac{Q^2 (1 - z_A)(1-z_B)}{1 - (1-z_A)(1 - z_B)} \, . \label{e.qtmax}
\end{equation}
Note that $\Tscsq{q}{}$ can be larger or smaller than 
$Q^2$ depending on $z_A$ and $z_B$. The invariant 
mass-squared of the dihadron pair is
\begin{equation}
\parz{p_A + p_B}^2 = z_A z_B \parz{Q^2 + \Tscsq{q}{}}\, ,
\end{equation}
which is of size $Q^2$ as long as $z_A$ and $z_B$ are fixed and not
too small. 

%-------------------------------------------------%
\subsection{The transverse momentum differential cross section}
\label{ss.xsec}

Written in terms of a leptonic and a hadronic tensor, the cross section
under consideration is
\begin{equation}
E_AE_B\frac{\diff{\sigma_{AB}}}{\diff{^3 \3{p}_A}{}\diff{^3 \3{p}_B}{}} = \frac{\alpha_{\rm em}^2}{8 \pi^3 Q^6} L_{\mu \nu} W^{\mu \nu} \label{e.tensors}
\end{equation}
where the leptonic tensor is 
\begin{equation}
L^{\mu \nu} \equiv l^\mu l'^{\nu} + l'^{\mu} l^\nu - g^{\mu \nu} l' \cdot l \, ,
\end{equation}
and the hadronic tensor is 
\begin{equation}
W^{\mu \nu} \equiv 4 \pi^3 \sum_{X} 
\langle 0 | j^{\mu}(0) |p_A, p_B, X \rangle  
\langle p_A, p_B, X | j^{\nu}(0) |0 \rangle 
\delta^{(4)}(q - p_A - p_B - p_X) \, , \label{e.hadten}
\end{equation}
where $j$ is the electromagnetic current, $p_X$ is the momentum of the unobserved part of the final state, and the $\sum_X$ includes all sums and integrals over unobserved final states $X$. 
The structure functions are related to the hadronic tensor through the
decomposition
\begin{align}
\label{e.hadsf}
W^{\mu\nu}(q,p_A,p_B)
=&\left(-g^{\mu\nu}+\frac{q^\mu{q}^\nu}{Q^2}-Z^\mu{Z}^\nu\right)W_T+Z^\mu{Z}^\nu{W}_L \, .
\end{align}
where $W_T$ and $W_L$ are the unpolarized structure functions. The $T$ and $L$ subscripts denote transverse and 
longitudinal polarizations respectively for the virtual 
photon.
For our purposes, we may neglect polarization and azimuthally dependent structure functions~\cite{Collins:2011qcdbook}. 
A convenient way to extract each structure function 
in \eref{hadsf} is to contract the hadronic tensor with associated extraction tensors,
$P_L^{\mu \nu}$ and $P_T^{\mu \nu}$:
\begin{equation}
W_T=P_T^{\mu\nu}W_{\mu\nu}\, , \qquad W_L=P_L^{\mu\nu}W_{\mu\nu}\, , \label{e.exttens}
\end{equation}
where  
\begin{equation}
\label{e.extens}
P_T^{\mu\nu}=\frac{1}{3}\left(-g^{\mu\nu}-Z^\mu{Z}^\nu+X^\mu{X}^\nu\right) \, , \qquad
P_L^{\mu\nu}=Z^\mu{Z}^\nu \, ,
\end{equation}
with the $Z^\mu$ and $X^\mu$ defined as in \eref{unitvects}.

After changing variables to $z_A$, $z_B$, $\Tsc{q}{}$ (see \aref{cov} for details),
\begin{equation}
\frac{\diff{\sigma_{AB}}}{\diff{z_A}\diff{z_B}\diff{\Tsc{q}{}}\diff{\cos\theta}\diff{\phi}}= 
\frac{\alpha_{{\rm em}}^2z_Az_B\parz{Q^2+\Tsc{q}{}^2}^2\Tsc{q}{}}{32\pi^2{Q}^6}\left[\left(1+\cos^2\theta\right)W_T+\sin^2\theta\ {W}_L \right] \, ,
\label{e.cross2}
\end{equation}
where $\theta$ and $\phi$ are the polar and azimuthal angles of 
lepton $l$ with respect to the $Z$ and $X$ directions in the photon 
frame. For the polarization independent case considered 
in this paper, we integrate this over $\theta$ and $\phi$ to get 
\begin{equation}
\frac{\diff{\sigma_{AB}}}{\diff{z_A}\diff{z_B}\diff{\Tsc{q}{}}}=\frac{\alpha_{{\rm em}}^2z_Az_B\parz{Q^2+\Tsc{q}{}^2}^2\Tsc{q}{}}{12\pi{Q}^6}\left[2W_T+W_L\right].
\label{e.cross3}
\end{equation}

In the small transverse momentum limit, the process in \eref{theprocess} is the one most simply and directly 
related to TMD ffs through 
derivations such as Ref.~\cite{Collins:1981uk} or 
more recently in Ref.~\cite[Chapt. 13]{Collins:2011qcdbook}. Note that, apart from the dihadron pair, the final state is totally inclusive (with no specification 
of physical jets or properties like thrust). This and 
the measurement of transverse momentum relative to a 
$Z$-axis as defined as above is important for the derivation of 
factorization, at least in its most basic form, with standard TMD and collinear ffs as the relevant correlation functions. Measurements within a jet and relative to a thrust axis~\cite{Seidl:2019jei} of course contain important information in relation to TMD ffs, but the connection is less direct.

%%%%%%%%%%%%%%%%%%%%%%%%%%%%%%%%%%%%%%%%%%%%%%%%%%%
\section{Factorization at Large, Moderate and Small Transverse Momentum}
\label{s.ptcalc}

To calculate in perturbative QCD, the differential cross section in \eref{cross3} needs to be factorized into a hard part and ffs, and 
different types of factorization are appropriate depending on the particular kinematical regime. Assuming $z_{A,B}$ are large enough to ensure that hadrons originate from separately fragmenting quarks, the three kinematical regions of interest for semi-inclusive
scattering are determined by the transverse momentum $\Tsc{q}{}$. There 
are three major regions: i.) $\Tsc{q}{} \sim Q$ so that $\Tsc{q}{}$ and $Q$ are equally viable hard scales, ii.) 
$m \ll \Tsc{q}{} \ll Q$ so that small $\Tsc{q}{}$ approximations are useful but $\Tsc{q}{}$ is large enough that intrinsic non-perturbative effects are negligible and logarithmic enhancements are only a small correction, iii.) $\Tsc{q}{} \lesssim m$ and all aspects of a TMD-based treatment are needed, including non-perturbative intrinsic transverse momentum (see also \sref{qthardness}). We will briefly summarize the calculation of each of these below. 

%-------------------------------------------------%
\subsection{The fixed $\order{\alpha_s}$ cross section at
large transverse momentum}
\label{ss.FO}

The scenario under consideration is one in which the two observed
hadrons are produced  at wide angle (so that $(p_A + p_B)^2 \sim
Q^2$), but are far from back-to-back (so that $\Tsc{q}{} \sim Q$).
This requires at least one extra gluon emission in the hard part. 
See \fref{siaA} (A) for the general structure of Feynman
graphs contributing at large $\Tsc{q}{}$ and for 
our momentum labeling
conventions. 
\begin{figure}
\centering
\begin{tabular}{c@{\hspace*{.5cm}}c@{\hspace*{.5cm}}}
\hspace*{-0.3cm}
\includegraphics[trim= 0in 2in 0in 0in,clip,width=0.5\textwidth]{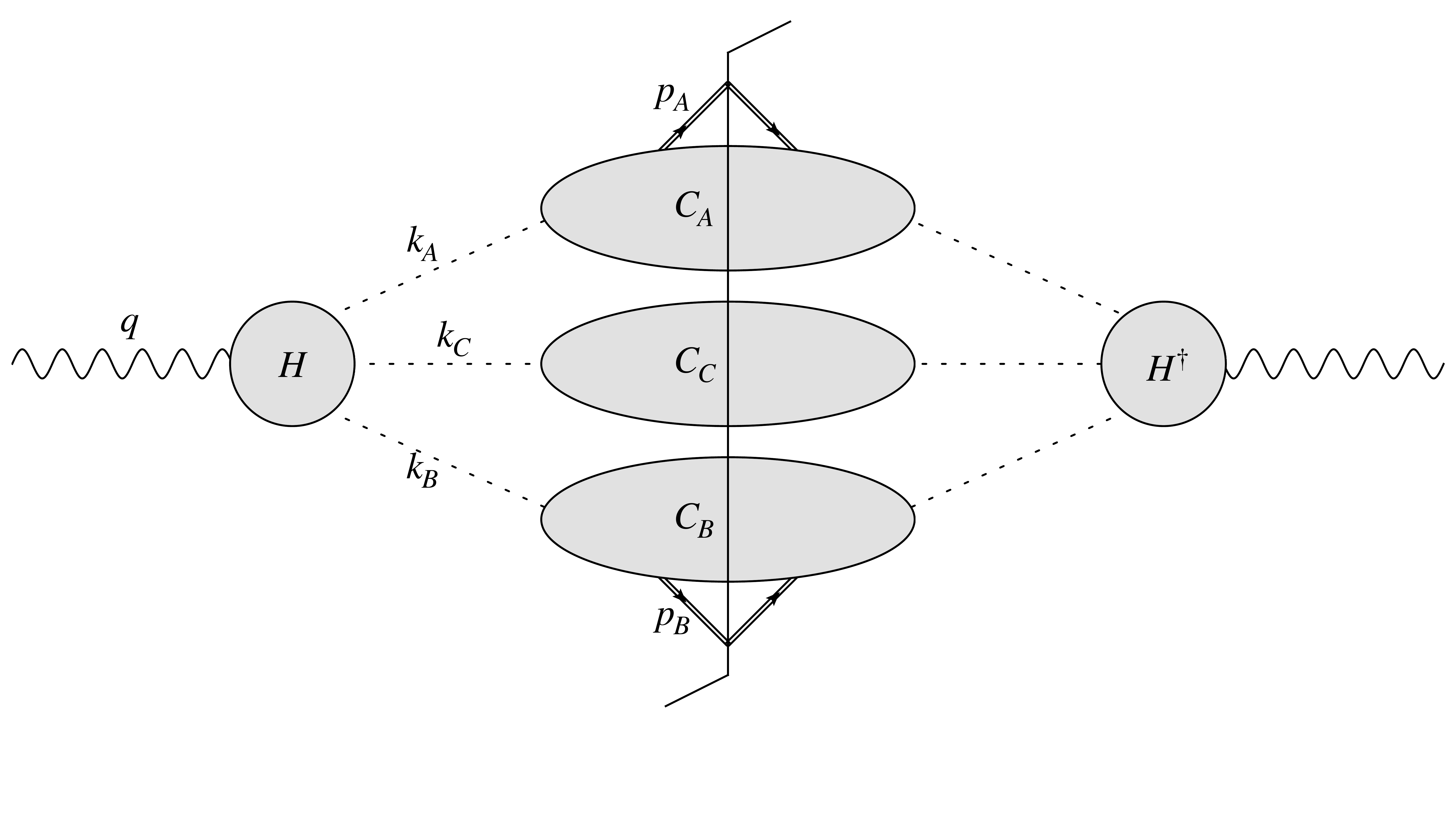}\ \
&
\includegraphics[width=0.5\textwidth]{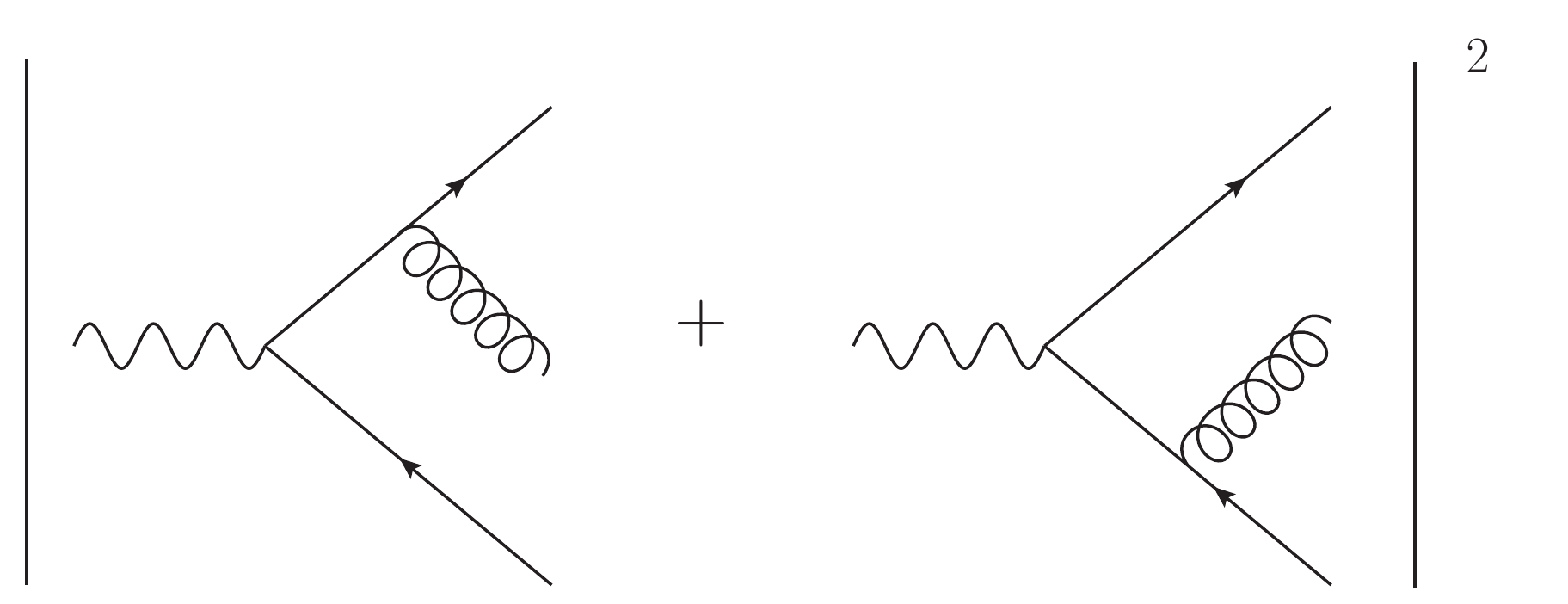}\\
(a) \ \ & (b) \vspace{10mm} \\ 
\end{tabular}
\caption{
  (a) The general diagrammatic structure contributing to
  \eref{theprocess} at large $\Tsc{q}{}$ and at LO in $\alpha_s$.  The
  outgoing partonic lines are dotted to 
  indicate that generally they can be of any type. In the region of interest for this paper, their momenta deviate by wide angles from the back-to-back orientation for the dihadron pair.
  $H$ represents the hard part of the interaction and the $C_{A,B,C}$ are the collinear subgraphs~\cite{Collins:2011qcdbook}. 
  (b) The $\order{\alpha_s}$ partonic contribution to the square-modulus 
  amplitude in the factorization of (a).}
\label{f.siaA}
\end{figure}

\begin{figure}
\centering
\begin{tabular}{c@{\hspace*{.5cm}}c@{\hspace*{.5cm}}c}
\hspace*{-0.3cm}
\includegraphics[width=0.26\textwidth]{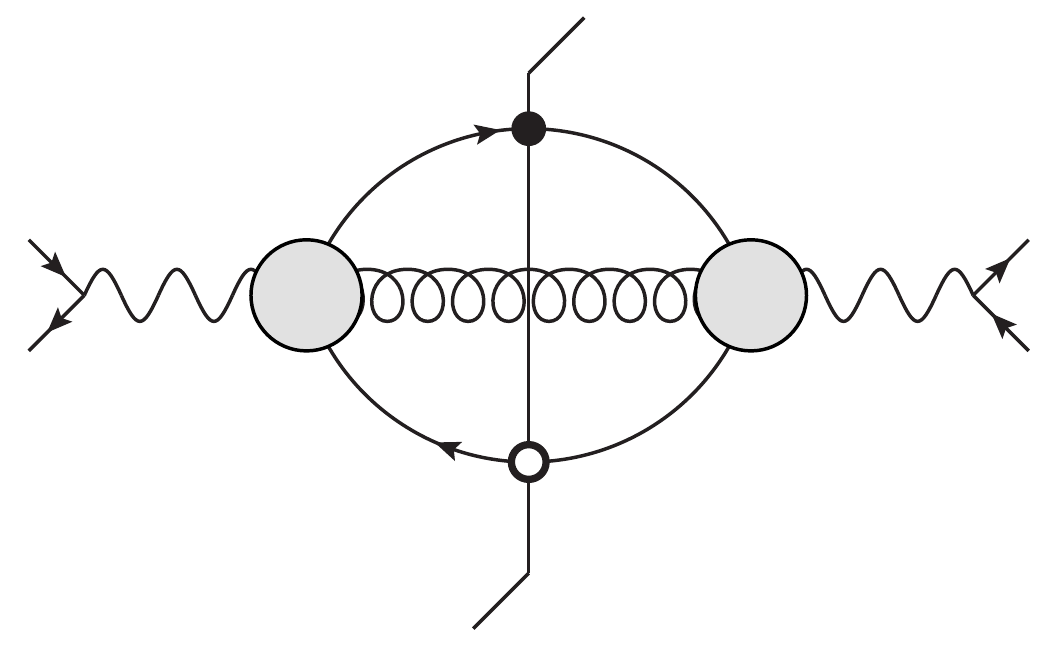}\ \
&
\includegraphics[width=0.26\textwidth]{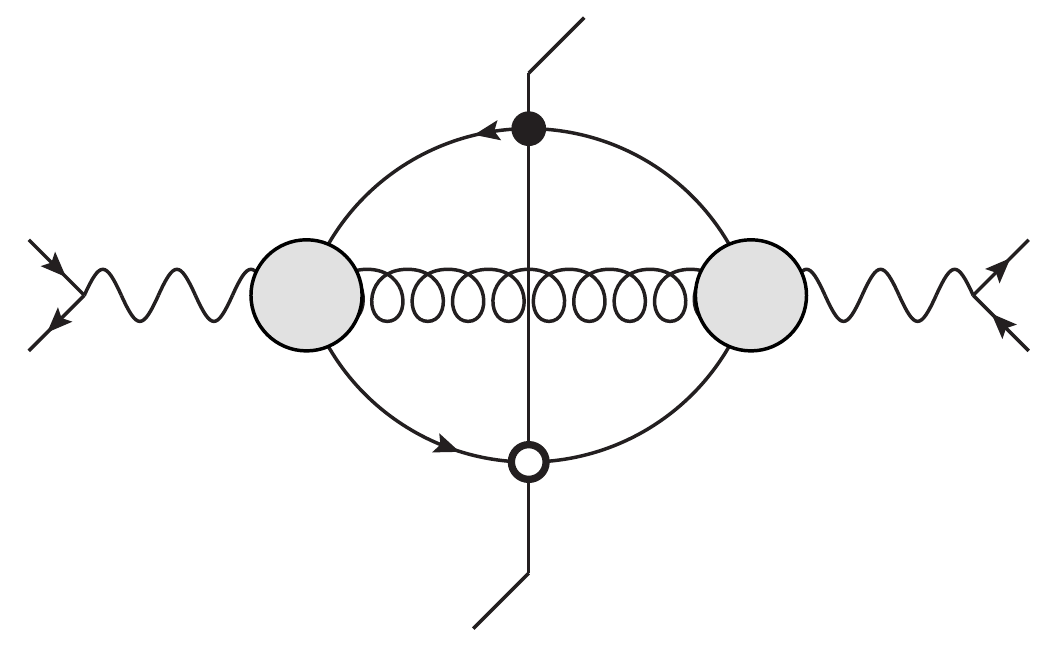}\ \
&
\includegraphics[width=0.26\textwidth]{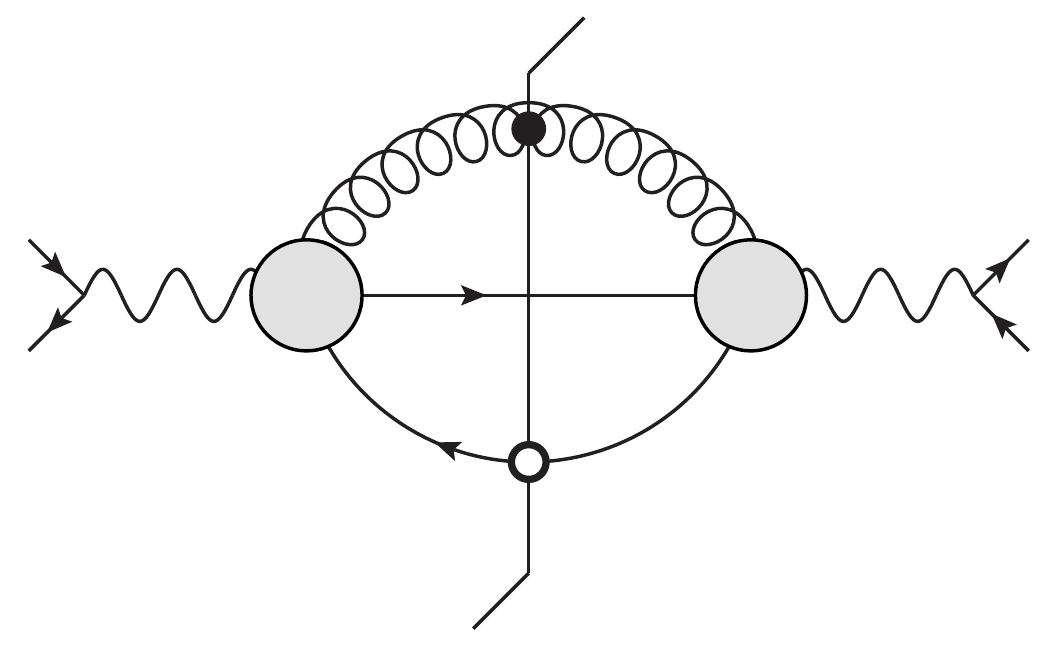}\\
(A) \ \ & (B) \ \ & (C)  \vspace{10mm} \\ 
\hspace*{-0.3cm}
\includegraphics[width=0.26\textwidth]{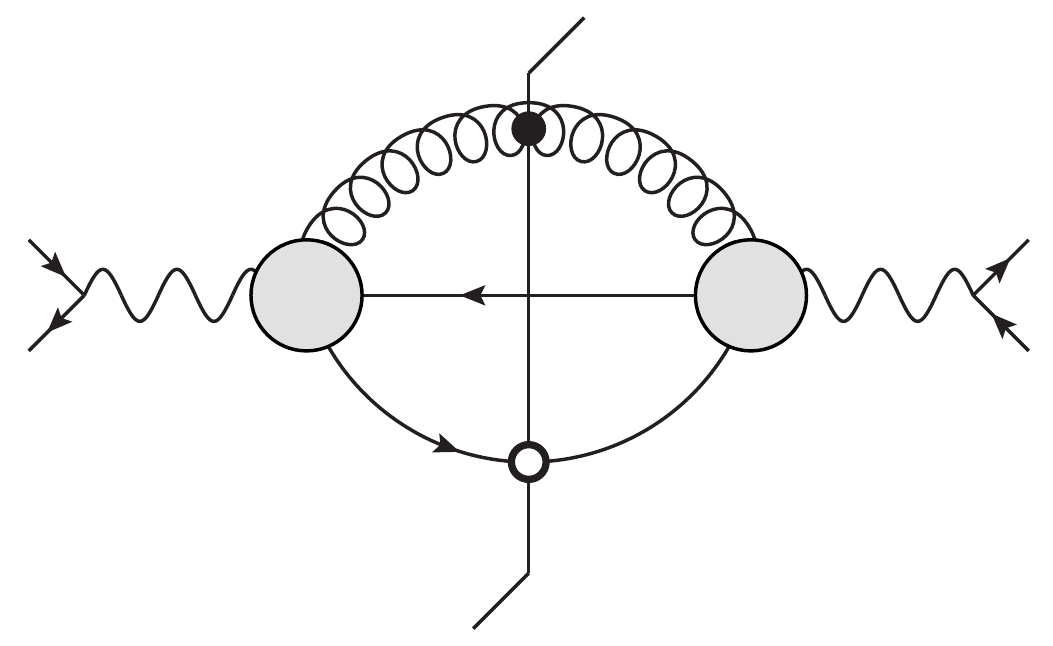}\ \
&
\includegraphics[width=0.26\textwidth]{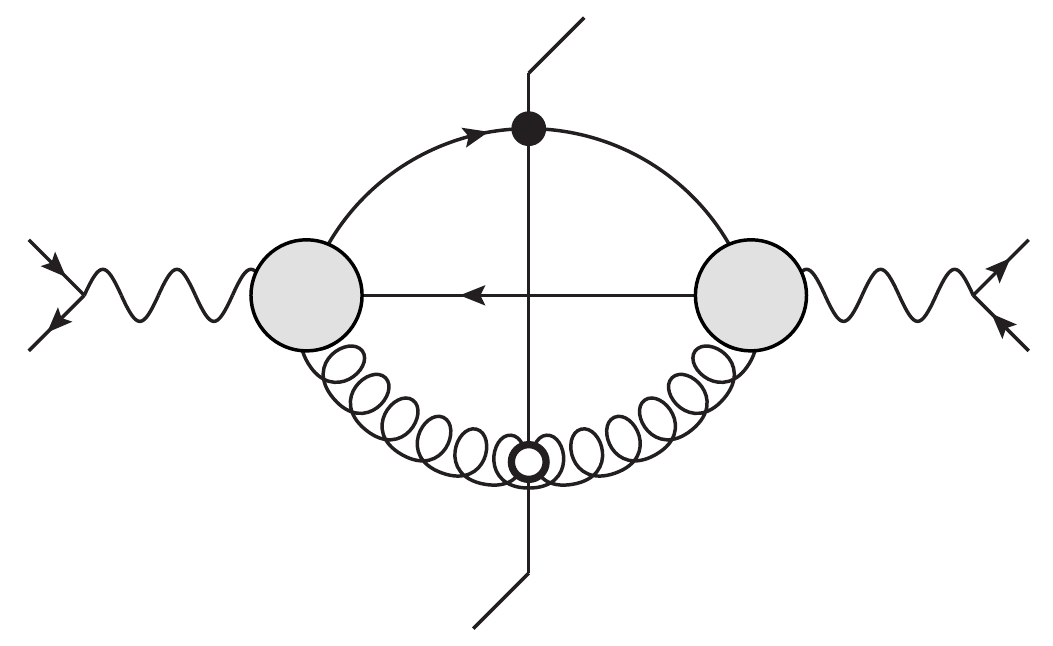}\ \
&
\includegraphics[width=0.26\textwidth]{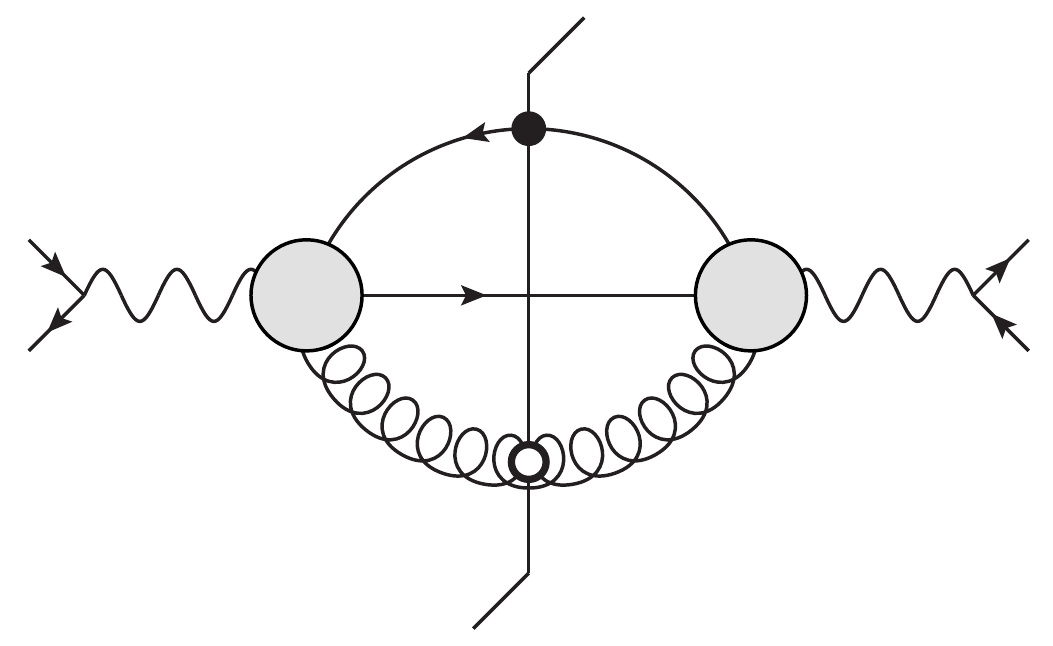}\\ 
(D) \ \ & (E) \ \ & (F) \vspace{10mm} 
\end{tabular}
\caption{Partonic channels that contribute at order $\alpha_s$. Detailed explanation in Sec.~\ref{ss.FO}.}
\label{f.graphs}
\end{figure}

The basic statement of collinear factorization for the differential
cross section is 
\begin{equation}
E_A E_B\frac{\diff{\sigma_{AB}}}{\diff{^3 \3{p}_A}{}\diff{^3 \3{p}_B}{}} 
= \sum_{i,j}\int_{z_A}^1\diff{\zeta_A}\int_{z_B}^1 
\diff{\zeta_B} 
\parz{E_AE_B\frac{\diff{\hat{\sigma}_{ij}}(\hat{z}_A,\hat{z}_B)}{\diff{^3 \3{p}_A}{}\diff{^3 \3{p}_B}{}}}
 d_{H_A/i}(\zeta_A)d_{H_B/j}(\zeta_B) \, 
\end{equation}
where the hat on the cross section in the integrand indicates that it
is for the partonic subprocess $l_1 + l_2 \to k_A + k_B + X$. $k_A$
and $k_B$ will label the momenta of the partons that hadronize.  The
integrals are over the momentum fraction variables $\zeta_A$ and
$\zeta_B$ that relate the hadron and parton momenta in
\fref{siaA}:
\begin{equation}
k_A \equiv p_A/\zeta_A \, , \qquad  
k_B \equiv p_B/\zeta_B \, . \label{e.momfracs}
\end{equation}
The $i,j$ sum is over the different possible flavors of parton that
can hadronize, $i,j \in \{u,d,g,\bar{u} \dots \}$. The number of active flavors depends on the scale. 
The $d_{H_A/i}(\zeta_A)$ and $d_{H_B/j}(\zeta_B)$ are the fragmentation
functions for flavor $i$($j$) partons to hadronize into hadrons of
flavor $A$ ($B$). We use the standard abbreviations 
\begin{equation}
\hat{z}_A = z_A/\zeta_A\, , \qquad \hat{z}_B = z_B/\zeta_B \, ,
\end{equation}
which follow from \eref{momfracs} and the partonic analogue of the
definitions in \eref{zdefs}.  
The momentum of the parton whose hadronization is unobserved is $k_C$~\cite{Collins:2007ph,Accardi:2008ne,Accardi:2019luo}. 
After factorization, the hard part involves the square-modulus of the
$H$ subgraph with massless, on-shell external partons. The graphs that
contribute to this at lowest order are shown in \fref{siaA}(b). 

It is useful to define a partonic version of the hadronic tensor,  
\begin{equation}
\widehat{W}_{ij}^{\mu \nu} \equiv 4 \pi^3 \sum_{X} 
\langle 0 | j_{ij}^{\mu}(0) |k_A, k_B, X \rangle  
\langle k_A, k_B, X | j_{ij}^{\nu}(0) |0 \rangle 
\delta^{(4)}(q - k_A - k_B - p_X) \, , \label{e.parten}
\end{equation}
in which case
\begin{equation}
 W^{\mu \nu} = \sum_{i,j}\int_{z_A}^1\frac{\diff{\zeta_A}}{\zeta_A^2}\int_{z_B}^1 
\frac{\diff{\zeta_B}}{\zeta_B^2} 
\widehat{W}_{ij}^{\mu \nu}(\hat{z}_A,\hat{z}_B)
 d_{H_A/i}(\zeta_A)d_{H_B/j}(\zeta_B) \, .
\end{equation}
Working with the hadronic tensor and with the extraction tensors like
\eref{exttens} conveniently automates the steps to obtain any
arbitrary structure function. The differential cross section is
\begin{align}
\label{e.xsec1}
\frac{\diff{\sigma_{AB}}}{\diff{z_A}\diff{z_B}\diff{\Tsc{q}{}}}=&\sum_{i,j}\int_{z_A}^1
\frac{\diff{\zeta_A}}{\zeta_A}\int_{z_B}^1\frac{\diff{\zeta_B}}{\zeta_B}
\parz{\frac{\diff{\hat{\sigma}}_{ij}(\hat{z}_A,\hat{z}_B)}{\diff{\hat{z}}_A\diff{\hat{z}}_B\diff{\Tsc{q}{}}}}d_{H_A/i}(\zeta_A)d_{H_B/j}(\zeta_B) \, ,
\end{align}
and the partonic cross section can be expressed analogously to
\eref{cross3},
\begin{equation}
\frac{\diff{\hat{\sigma}}_{ij}}{\diff{\hat{z}}_A\diff{\hat{z}}_B\diff{\Tsc{q}{}}}
=\frac{\alpha_{\rm em}^2\hat{z}_A\hat{z}_B\parz{Q^2+\Tsc{q}{}^2}^2\Tsc{q}{}}{12\pi{Q}^6}\left[2\widehat{W}_{T,ij}+\widehat{W}_{L,ij}\right] \, ,
\end{equation}  
where $\widehat{W}_{T,ij}$ and $\widehat{W}_{L,ij}$ are partonic
structure functions calculated from the graphs in \fref{siaA}(b). 

Given 
the expressions for the squared amplitudes in \fref{siaA}(b), the
evaluation of the differential cross section becomes straightforward.
Each possible combination of final state parton pairs in
\fref{siaA}(b) can hadronize into $H_A$ and $H_B$ with fragmentation
functions that depend on both the fragmenting parton and final state
hadron. Six such channels contribute at leading order in $\alpha_s$, and we
organize these diagrammatically in \fref{graphs}, with $k_A$, $k_B$
and $k_C$ assigned to the quark, antiquark or gluon according to whether it hadronizes to $H_A$, $H_B$, or is unobserved. 
A solid dot marks the parton that hadronizes into $H_A$ (always $k_A$
parton momentum) and the open dot marks the parton that hadronizes
into $H_B$ (always $k_B$ momentum). There is an integral over all momentum of the remaining line
($k_C$). Quark lines include all active quark
flavors, and are shown separately from the anti-quark lines since they
correspond to separate ffs. Notice that, unlike in the case of the $q_T$-integrated cross section for single hadron production, there is
already sensitivity to the gluon fragmentation function at the lowest
non-vanishing order. 
The analytic expressions needed for the
calculation are summarized in \aref{fo}.

%-------------------------------------------------%
\subsection{The asymptotic $\frac{\Tscsq{q}{}}{Q^2} \to 0$ limit}
\label{ss.asy}

The small $\Tscsq{q}{}/Q^2$ limit of \eref{xsec1} involves
considerable simplifications analogous to those obtained in TMD
factorization, but applied to fixed order massless partonic graphs. It is potentially a useful simplification, therefore, in situations 
where $\Tscsq{q}{}$ is small enough that a $\Tscsq{q}{}/Q^2$ expansion applies, but still large enough that fixed order perturbative calculations are reasonable approximations. As we will see in later sections, it is also useful for estimating the borders of the regions 
where small $\Tscsq{q}{}/Q^2$ approximations are appropriate.
 
The asymptotic term is obtainable by directly expanding the fixed
order calculation in powers of small $\Tsc{q}{}/Q$, with a careful
treatment of the soft gluon region in the integrals over $\zeta_A$ and
$\zeta_B$. The steps are similar to those in SIDIS, and we refer
to Ref.~\cite{Nadolsky:1999kb} for a useful discussion of them.  When
performed for the $e^+ e^-$ annihilation case under consideration
here, the result is
\begin{align}
\label{e.asyxsec}
\frac{\diff{\sigma_{AB}^{ASY}}}{\diff{z_A}\diff{z_B}\diff{\Tsc{q}{}}}=\frac{4 \alpha_{\rm em}^2 \alpha_s}{Q^2 \Tsc{q}{}}\sum_{q} e_q^2&\left\{2C_F\left[\ln{\parz{\frac{Q^2}{\Tscsq{q}{}}}}-\frac{3}{2}\right]\parz{d_{H_A/q}(z_A)d_{H_B/\bar{q}}(z_B)+d_{H_A/\bar{q}}(z_A)d_{H_B/q}(z_B)}\right.\no
&+d_{H_A/q}(z_A)\left[(P_{\bar{q}\bar{q}}\otimes{d}_{H_B/\bar{q}})(z_B)+(P_{g\bar{q}}\otimes{d}_{H_B/g})(z_B)\right]\bigg.\no
&+d_{H_A/\bar{q}}(z_A)\left[(P_{qq}\otimes{d}_{H_B/q})(z_B)+(P_{gq}\otimes{d}_{H_B/g})(z_B)\right]\bigg.\no
&+d_{H_B/q}(z_B)\left[(P_{\bar{q}\bar{q}}\otimes{d}_{H_A/\bar{q}})(z_A)+(P_{g\bar{q}}\otimes{d}_{H_A/g})(z_A)\right]\bigg.\no
&+d_{H_B/\bar{q}}(z_B)\left[(P_{qq}\otimes{d}_{H_A/q})(z_A)+(P_{gq}\otimes{d}_{H_A/g})(z_A)\right]\bigg\} \, ,
\end{align}
where $P_{ij}$ are the leading order unpolarized splitting functions
\begin{equation}
P_{qq}(z)=P_{\bar{q}\bar{q}}(z)=C_F\left[\frac{1+z^2}{\parz{1-z}_+}+\frac{3}{2}\delta\parz{1-z}\right] \, , \qquad 
P_{gq}(z)=P_{g\bar{q}}(z)=C_F\left[\frac{1+\parz{1-z}^2}{z}\right] \, ,  \label{e.pdists}
\end{equation}
and $\otimes$ represents the convolution integral
\begin{equation}
(f\otimes{g})(z)=\int_{z}^1\frac{\diff\zeta}{\zeta}f(z/\zeta)g(\zeta) \, .
\end{equation}
The ``$()_+$" in \eref{pdists} denotes the usual 
plus-distribution. The ``$\rm ASY$'' superscript on 
\eref{asyxsec} symbolizes the asymptotically small 
$\Tscsq{q}{}/Q^2$ limit for the cross section. The sum over $q$ is a sum over all 
active quark flavors. 

%--------------------------------------------------------
\subsection{TMD ffs and the small $q_T$ region }
\label{ss.smallqt}

In the small transverse momentum limit of the 
cross section, the $W_L$ structure function becomes 
power suppressed. The cross section in \eref{cross3} is 
simply 
\begin{equation}
\frac{\diff{\sigma_{AB}}}{\diff{z_A}\diff{z_B}\diff{\Tsc{q}{}}}=\frac{\alpha_{{\rm em}}^2z_Az_B \Tsc{q}{}}{6\pi{Q}^2} W_T \, ,
\label{e.cross3small}
\end{equation}
and the structure function $W_T$ (or hadronic tensor) factorizes in a well known way into TMD fragmentation functions
\begin{align}
W_T = \frac{8\pi^3 z_A z_B}{Q^2}\sum_q  \widehat{W}_{T,q}
      \int \frac{\diff{^{2}\T{b}{}}}{(2 \pi)^{2} }
      e^{-i \T{b}{} \cdot \T{q}{}}
      \left[
      \tilde{D}_{H_A/q} \tilde{D}_{H_B/\bar{q}}
            +\tilde{D}_{A/\bar{q}} \tilde{D}_{B/q}
      \right] \, , 
\end{align}
where 
\begin{equation}
\widehat{W}_{T,q} = 6  Q^2 e_q^2 \, .
\end{equation}
The $\tilde{D}_{H/q}$ are the TMD fragmentation 
functions in transverse coordinate $\bm{b}_T$ space. After evolution, the TMD ff for a hadron $H$ from quark 
$q$ is
\begin{align}
\tilde{D}_{H/q}&(z,\bm{b}_T;\mu,\zeta_D)
=\sum_j \int_z^1 \frac{d\hat{z}}{\hat{z}^{3}} 
   \tilde{C}_{j/q}(z/\hat{z},b_*;\zeta_D,\mu) d_{H/j}(\hat{z},\mu_b)
\notag\\
&\times\exp\left\{\ln\frac{\sqrt{\zeta_D}}{\mu_b}\tilde{K}(b_*;\mu_b)    
+\int_{\mu_b}^{\mu}\frac{d\mu'}{\mu}
    \left[
    \gamma(\mu';1) - \ln\frac{\sqrt{\zeta_D}}{\mu'} \gamma_K(\mu')
    \right]
+g_{H/j}(z,b_T) + \frac{1}{2}g_K(b_T) \ln\frac{\zeta_D}{\zeta_{D,0}}
\right\}
\label{e.tmd}
\end{align}
The $j$ index runs over all quark flavors and includes gluons, and the functions $d_{H/j}(z,\mu_b)$ are ordinary collinear ffs which are convoluted
with coefficient functions $C_{j/q}$ derived from the the small $b_T$ limit of the TMDs.  All perturbative contributions, $C_{j/q}$, $\tilde{K}$, $\gamma$, and $\gamma_K$ are known by now to several orders in $\alpha_s$~\cite{Rogers:2015sqa,Scimemi:2019mlf}.  
However, non-perturbative functions also enter to parametrize the truly non-perturbative and intrinsic parts of the TMD functions. 
These are $g_{H/j}$, which is hadron and flavor dependent, and $g_K$, which is  
independent of the nature of hadrons and parton flavors and controls the non-perturbative contribution to the 
evolution.  When combined in a cross section $\zeta_{D_A} \times \zeta_{D_B} = Q^4$. Some common parametrizations used for phenomenological fits are
\begin{align}
g_{H/j}(z,b_T) &=-\frac{1}{4 z^2} \langle \Tscsq{K}{H/j,} \rangle b_T^2\, , \label{e.g1} \\
g_K(b_T)       &=-\frac{1}{2}g_2 b_T^2 \, . \label{e.g2}
\end{align}
Perturbative parts of calculations are usually regulated in the large $b_T$ region by using, for example, the $b_*$ prescription with:
\begin{align}
b_*(b_T)=\frac{b_T}{\sqrt{1+\left(b_T/b_{\rm max}\right)^2}},\hspace{1cm}
\mu_b(b_*)\propto\frac{1}{b_*} \, .
\end{align}
While there are many ways to regulate large $b_T$, and many alternative proposals for parametrizing the non-perturbative TMD inputs $\langle \Tscsq{K}{H/j,} \rangle$ and $g_2$, the above will be sufficient for the purpose of capturing general trends in the comparison of large and small transverse momentum calculations in \sref{analysis}.  

%%%%%%%%%%%%%%%%%%%%%%%%%%%%%%%%%%%%%%%%%%%%%%%%%%%
\section{Transverse Momentum Hardness}
\label{s.qthardness}

The question of what constitutes large or small transverse momentum
warrants special attention, so  we now consider how the kinematical
configuration of the third parton in graphs of the form of
\fref{siaA}(a), not associated with a fragmentation function, affects
the sequence of approximations needed to obtain various types of
factorization.\footnote{For this section we allow for the possibility
of arbitrarily many hard loops inside $H$.} Generally, the propagator
denominators in the hard blob $H$ can be classified into two types
depending on whether  $k_C$ attaches inside a far off-shell virtual
loop or to an external leg.  If it attaches inside a virtual loop, the
power counting is 
\begin{equation}
\frac{1}{2\ k_C \cdot k_{A,B} + O(Q^2)} \, , \label{e.prop1}
\end{equation}
and for an external leg attachment (the off-shell propagators in
\fref{siaA}(b), for example)
\begin{equation}
\frac{1}{2\ k_C \cdot k_{A,B} +  O(m^2)} \, . \label{e.prop2}
\end{equation}
The coefficients of the $O(Q^2)$ and $O(m^2)$ are numerical factors
roughly of size $1$. Here the $m^2$ is a small mass scale comparable
to $\Lambda_{\rm QCD}^2$ or a small hadron mass-squared.  Possible
$\order{m^2}$ terms in the \eref{prop1} denominator can always be
neglected relative to $\order{Q^2}$ and so have not been written
explicitly.

The question that needs to be answered to justify collinear versus TMD
factorization is whether the $2\ k_C \cdot k_{A,B}$ terms are also
small enough to be dropped, or if they are large enough that they can
be treated as hard scales comparable to $Q^2$, or if the true
situation is somewhere in between.  The fixed order calculations like
those of the previous section is justified if 
\begin{equation}
\left| \frac{2\ k_C \cdot k_{A,B}}{Q^2} \right|
\end{equation}
is not much smaller than $1$. A quick estimate of the relationship
between this ratio and $\Tscsq{q}{}/Q^2$ is obtained as follows:
\begin{equation}
\left| \frac{2\ k_C \cdot k_{A,B}}{Q^2} \right| \approx \left| \frac{(q - k_{B,A})^2}{Q^2} \right| \approx \left| \frac{(q - \frac{p_{B,A}}{z_{B,A}})^2}{Q^2} \right| = \frac{\Tscsq{q}{}}{Q^2} \, , \label{e.hardness}
\end{equation}
where the first ``$\approx$'' means momentum conservation is used
with $k_{A,B,C}^2 \approx 0$, and the second ``$\approx$'' means the
standard small $\Tscsq{q}{}$ approximation  for the photon vertex,
$\zeta_A \approx z_A$, is being used. For the denominator in
\eref{prop2}, the relevant ratio is $m^2/ (2\ k_C \cdot k_{A,B})$, and
arguments similar to the above give 
\begin{equation}
\left| \frac{m^2}{2\ k_C \cdot k_{A,B}} \right| \approx \frac{m^2}{\Tscsq{q}{}}  \label{e.hardness2} \, .
\end{equation}
If \eref{hardness} is $\order{1}$ while \eref{hardness2} is much less
than one, then the approximations on which collinear factorization at
large $\Tscsq{q}{}$ is based are justified. 

The situation is reversed if \eref{hardness2} is $\order{1}$ or larger
but \eref{hardness} is small. In that case, the neglect of the
$\order{m^2}$ effects (including intrinsic transverse momentum) in the
\eref{prop2} denominators is unjustified. However, the smallness of
\eref{hardness} means  neglecting the $2 k_C \cdot k_{A,B}$ terms in
the hard vertex is now valid, and this leads to its own set of extra
simplifications.  Ultimately, such approximations are analogous to
those used in the derivation of TMD factorization. 

An additional way to estimate the hardness of $\Tscsq{q}{}$ is to
compare with the kinematical maximum in \eref{qtmax}. For $z_{A,B}
\gtrsim .4$, it can produce a significantly smaller ratio than
\eref{hardness}. For example, for $z_{A,B} = .5$, $q_{\rm T}^{\rm Max}/Q^2 = 1/3$. Certainly, small $\Tscsq{q}{}/Q^2$ approximations
fail near such thresholds. 

The range of possible transverse momentum regions can be summarized
with three categories:
\begin{itemize}
\item \underline{Intrinsic transverse momentum:} \eref{hardness2} is
      of size $1$ or larger, but \eref{hardness} is a small suppression
      factor. TMD factorization, or a similar approach that accounts for
      small transverse momentum effects, is needed. Such a kinematical
      regime is ideal to studying intrinsic transverse momentum properties of
      fragmentation functions. 
\item \underline{Hard transverse momentum:} \eref{hardness2} is much
      less than $1$, and \eref{hardness} is comparable to $1$.  Therefore,
      fixed order calculations like those of the previous section are
      justified.
\item \underline{Intermediate transverse momentum:} \eref{hardness2}
      is much less than $1$, but \eref{hardness} is also much less than one.
      In this case, the previous two types of approximations are
      simultaneously justifiable.  Transverse momentum dependence is mostly
      perturbative, but large logarithms of $\Tscsq{q}{}/Q^2$ imply that
      transverse momentum resummation and/or TMD evolution are nevertheless
      important. 
\end{itemize}

The large transverse momentum fixed order calculations are the most
basic of these, since they involve only collinear factorization
starting with tree level graphs, so it is worthwhile to confirm that
there is a region where they are phenomenologically accurate, as is
the aim of the present paper.  Direct comparisons between fixed order
calculations and measurements can help to confirm or challenge the
above expectations. For example, consider a case 
where $Q \sim 10$~GeV while the largest measurable 
transverse momenta about $\sim 7$~GeV. 
 Then logarithms of $\Tscsq{q}{}/Q^2$, i.e., $| \ln
.7^2 | \sim .7$, are not large while \eref{hardness} is a
non-negligible $\sim 0.5$.  These are ideal kinematics, therefore, for
testing the regime where fixed order calculations are expected to
apply.

%%%%%%%%%%%%%%%%%%%%%%%%%%%%%%%%%%%%%%%%%%%%%%%%%%%
\section{Large and Small Transverse Momentum Comparison}
\label{s.analysis}
We begin our comparison by computing the fixed order collinear factorization based cross section for the $\Tscsq{q}{} \sim Q^2$ region using the DSS14 ff parametrizations~\cite{deFlorian:2014xna}, and we compare with the calculation of the asymptotic term in \eref{asyxsec}.
The results are shown for both moderate $Q\sim 12$~GeV and for large
$Q\sim 50$~GeV in \fref{matching} (left panel), with $z_{A,B} =
0.3$ in both cases.  The horizontal axis is the ratio
$\Tsc{q}{}/q_{\rm T}^{\rm Max}$, using \eref{qtmax} to make the
proximity to the kinematical large-$\Tscsq{q}{}$ 
threshold clearly visible. 

The exact kinematical relation (for $1 \to 3$ scattering) between $\zeta_B$ and $\zeta_A$ is
\begin{equation}
\label{e.zetAzetB}
\zeta_B = z_B 
\frac{(Q^2 + \Tscsq{q}{}) (z_A - \zeta_A)}{\Tscsq{q}{} z_A 
+ Q^2 (z_A - \zeta_A)} \, ,
\end{equation}
while the cross section in the asymptotically small 
$\Tscsq{q}{}/Q^2$ limit has either $\zeta_A = z_A$ with 
$\zeta_B \geq z_B$ or $\zeta_B = z_B$ with 
$\zeta_A \geq z_A$.  The asymptotic phase space in 
the $\zeta_B$-$\zeta_A$ plane 
approaches a rectangular wedge shape in the 
small $\Tscsq{q}{}$ limit, shown as the solid black lines
in \fref{matching} (right panel) for fixed values of $z_A = z_B$.
For comparison, the differently colored dashed, dot-dashed, and dotted lines show 
the 
$\zeta_B$-$\zeta_A$ curves from \eref{zetAzetB}
for 
various nonzero $\Tscsq{q}{}$. The deviation between the 
colored and black curves gives one indication of the 
degree of error introduced by taking the small 
$\Tscsq{q}{}$ limit. \fref{matching}(right panel) shows how these grow at large $z_{A,B}$. A non-trivial kinematical correlation forms between momentum fractions $\zeta_A$ and $\zeta_B$ in the large $z_A,z_B$ and large $\Tscsq{q}{}$ regions. 
Notice also that the contours are scale independent, since $q_{\rm T}^{\rm Max}$ is proportional to $Q^2$, so kinematical errors from small $\Tsc{q}{}$ approximations are likewise scale independent.

The point along the horizontal axis where the asymptotic term turns negative is
another approximate indication of the region above which small $\Tscsq{q}{}/Q^2$ approximations
begin to fail and the fixed order collinear factorization
 treatment should become more reliable, provided 
 $z_{A,B}$ are at fixed moderate values and $q_{\rm T}$ 
 is not too close to the overall kinematical thresholds.
 That point is shown in \fref{matching}(left) for two representative values of small ($Q = 12$~GeV) and large 
 $Q = 50$~GeV. 
 The  transition is at rather small transverse momentum, roughly $ q_{\rm T}/q_{\rm T}^{\rm Max}\sim
0.2$, though the exact position depends on a number of details, including the shapes of the collinear fragmentation functions. If the asymptotic term is used as the indicator, then the transition is also roughly independent of $Q$.

We are ultimately interested in asking how the fixed order 
collinear calculation compares with existing TMD ff parametrizations near the small-to-large transverse momentum transition point. 
A reasonable range of non-perturbative parameters like $\langle \Tscsq{K}{H/j,} \rangle$ and $g_2$ in \erefs{g1}{g2}, 
can be estimated from a survey of existing phenomenological fits. We will make the approximation that all light flavors have equal $\langle \Tscsq{K}{H/j,} \rangle = \langle \Tscsq{K}{} \rangle$ for pion production. Then
values for 
$\langle \Tscsq{K}{H/j,} \rangle$ lie in the range from about
$.11$~GeV$^{-2}$ to $.23$~GeV$^{-2}$~\cite{Bacchetta:2017gcc}, which straddles the value 
$0.16$~GeV$^{-2}$ in Ref.~\cite{Schweitzer:2010tt}. For $g_2$, we use a minimum value of 
$0$ to estimate the effect of having no non-perturbative evolution at all, and we use a maximum value of $.184$~GeV$^{-2}$, from Ref.~\cite{Konychev:2005iy}, which is 
at the larger range of values that have been extracted. This range also straddles the 
$g_2 = .13$~GeV$^{-2}$ found in Ref.~\cite{Bacchetta:2017gcc}. 
In all cases, we use the lowest order perturbative anomalous dimensions since these were used in most of the Gaussian-based fits above. 
Collectively, the numbers above produce the blue 
bands in \fref{pythia_cf} (left). The references quoted above generally include uncertainties for their parametrizations of $\langle \Tscsq{K}{j,} \rangle$ and $g_2$, but these are much 
smaller than the uncertainty represented by the blue band in \fref{pythia_cf} (left). 
We use a representative estimate of $b_{\rm max} = 1.0$~GeV$^{-1}$; Refs.~\cite{Bacchetta:2017gcc} and 
~\cite{Konychev:2005iy} use slightly larger values 
($1.123$~GeV$^{-1}$ and $1.5$~GeV$^{-1}$ respectively), but larger 
$b_{\rm max} \gtrsim 1.0$~GeV$^{-1}$ also has a small effect and only increases the general disagreement with the collinear fixed order calculation.
%%%%%%%%%%%%%%%%%%%%%%%%%
\begin{figure}
\centering
\includegraphics[width=\textwidth]{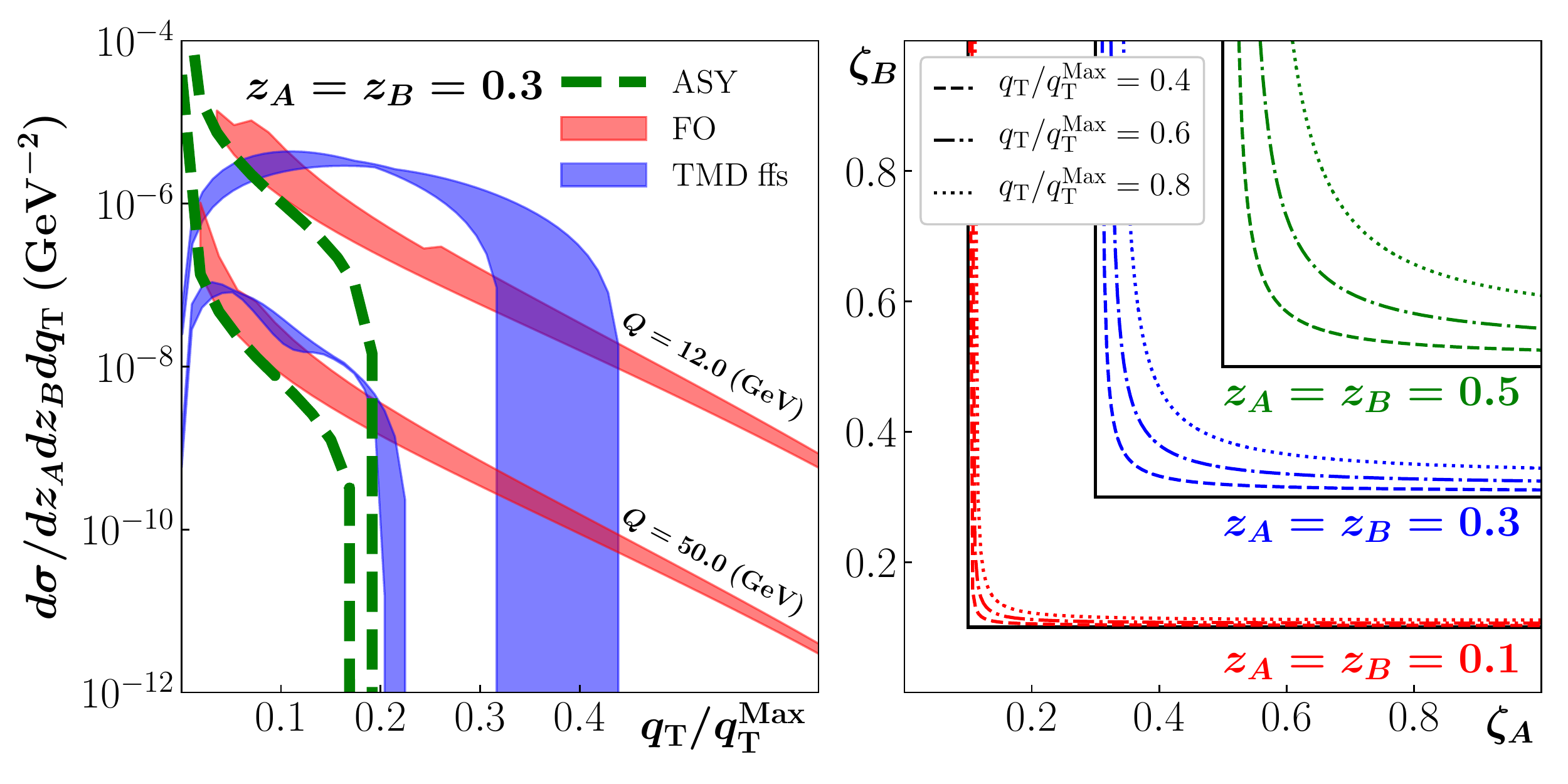}
\caption{(left): LO collinear factorization predictions for the inclusive $e^+ e^-$ to dihadron cross section (\sref{ptcalc} and \aref{fo}), for $Q=12,50$~GeV. The red 
band shows the range covered by switching the renormalization 
group scale between $\mu = Q$ (lower edge) and $\Tsc{q}{}$ (upper edge). The blue band is the calculation performed using TMD ffs, and the band shows the range covered by the 
values of the non-perturbative parameters discussed in Sec.~\ref{s.analysis}. 
(right): correlation between partonic momentum fractions $\zeta_{A,B}$ for various values of $q_{\rm T}/q_{\rm T}^{\rm Max}$.  
}
\label{f.matching}
\end{figure}
%%%%%%%%%%%%%%%%%%%%%%%%%
%%%%%%%%%%%%%%%%%%%%%%%%%
\begin{figure}
\centering
\includegraphics[width=\textwidth]{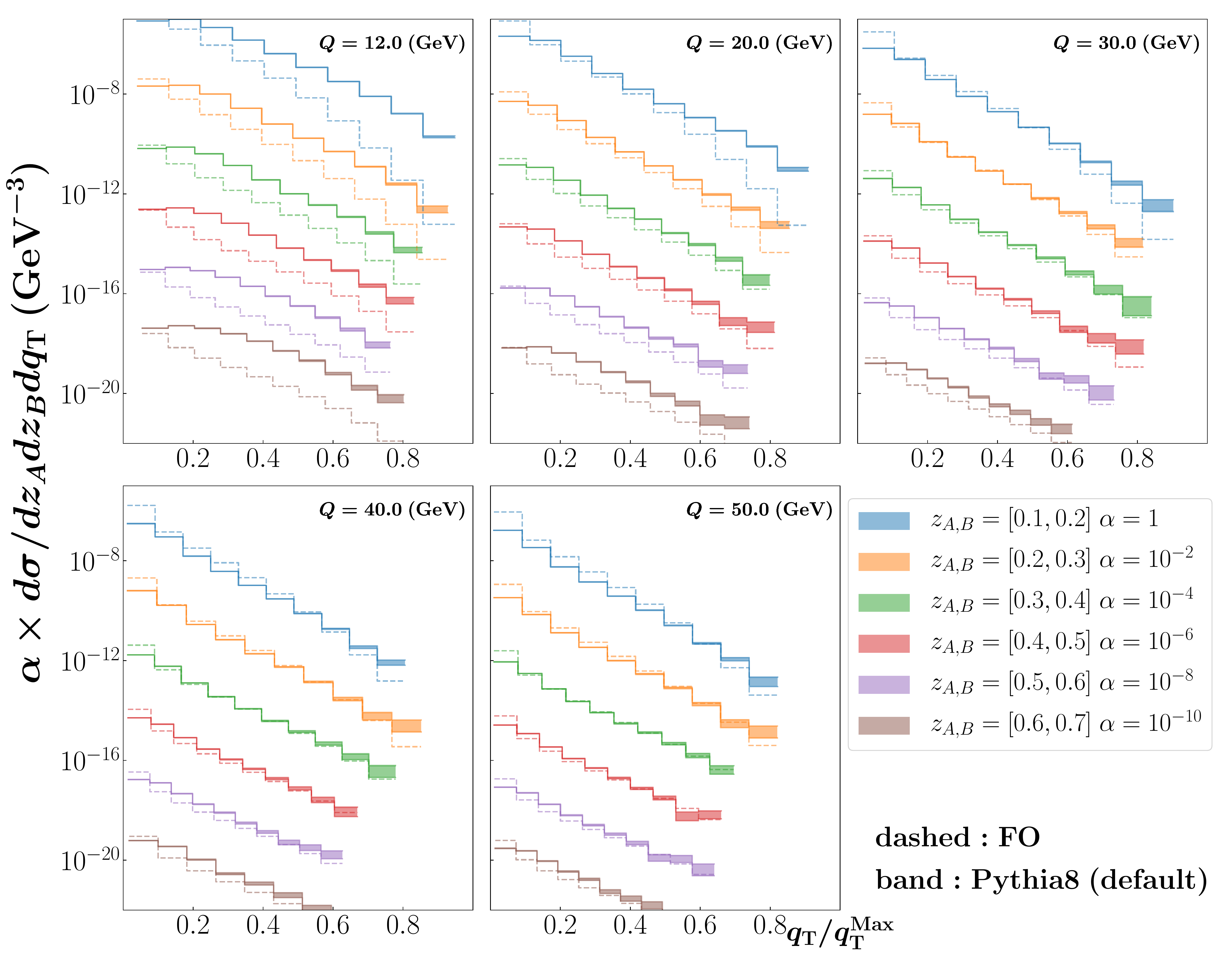}
\caption{The lowest order collinear factorization calculation from \sref{ptcalc} compared with $\pi^+/\pi^-$ pair production simulated by PYTHIA-8 with default settings for different ranges of $z_{A,B}$ and for increasing values of $Q$, starting with $Q = 12$~GeV. Both the fixed-order calculation and the simulation are averaged in the $z_{A,B}$ bins. The uncertainty on the bands is purely statistical. 
}
\label{f.pythia_cf}
\end{figure}
%%%%%%%%%%%%%%%%%%%%%%%%%%%%%%%%

Observe in \fref{matching} (left) that, despite our somewhat overly liberal band sizes for the 
TMD ff calculation, large tension in the intermediate transverse momentum region between the TMD ff-based cross section and the fixed order 
collinear calculation nevertheless remains. For the 
$z_{A,B} \approx .3$ shown, $q_{\rm T}^{\rm Max} \approx Q$.  The $Q= 50$~GeV curves show that 
as $Q$ is raised, this tension diminishes, though at a perhaps surprisingly slow rate. For $Q = 12$~GeV, the 
asymptotic and fixed order terms approach one another, but only at very small $\Tsc{q}{}$. The curves contained within the blue band deviate qualitatively from the asymptotic and fixed order terms across all transverse momentum, and the blue band badly overshoots both in the intermediate region of $\Tsc{q}{} \approx 2-3$ GeVs. The result is reminiscent of the situation with other processes -- see, for example, Fig.~6 of \cite{,Boglione:2014oea} for SIDIS.

Interestingly, data for the observable
of \eref{theprocess} for  
$\pi^+/\pi^-$ production simulated 
with PYTHIA 8~\cite{Sjostrand:2006za,Sjostrand:2014zea} using default settings, shows quite reasonable agreement with the collinear factorization calculation in the expected range of intermediate transverse momentum and $z_{A,B}$ and very large $Q$, validating the analytic fixed order collinear calculation in regions where it is most expected that the collinear calculations and the simulation should overlap.
We illustrate this in \fref{pythia_cf}, where for
$z_{A,B}$ between $0.2$ and $0.6$ the fixed order analytic calculation 
agrees within roughly a factor of 2 with the PYTHIA-generated spectrum for $Q \gtrsim 20$~GeV and for 
$ q_{\rm T}/q_{\rm T}^{\rm Max} \sim 0.5$. At smaller $Q \lesssim 20$~GeV, the agreement between the fixed order calculation and the simulation is much worse, though because $Q$ is relatively small and the event generator includes only the leading order hard scattering (with parton showering), it is unclear how the disagreement in that region should be interpreted. Nevertheless, it is interesting to observe that the trend wherein the collinear factorization calculation undershoots data, seen in SIDIS~\cite{Gonzalez-Hernandez:2018ipj} and Drell-Yan~\cite{Bacchetta:2019tcu} calculations, seems to persist even here. In the future, it would be interesting to perform a more detailed Monte Carlo study that incorporates treatments of higher order hard scattering.

%%%%%%%%%%%%%%%%%%%%%%%%%%%%%%%%%%%%%%%%%%%%%%%%%%%
\section{Conclusions}
\label{s.conclusion}
As one of the simplest processes with non-trivial transverse momentum dependence, 
dihadron production in $e^+ e^-$ annihilation is ideal for testing theoretical treatments of 
transverse momentum distributions generally. A goal of this paper has been to spotlight its possible use as a probe of the transition between 
kinematical regions corresponding to different types of QCD factorization. There have been a number of studies highlighting tension between large transverse momentum collinear factorization based calculation and cross section measurements for 
Drell-Yan and SIDIS, 
Refs.~\cite{Daleo:2004pn,Kniehl:2004hf,Gonzalez-Hernandez:2018ipj,Wang:2019bvb,Bacchetta:2019tcu}. Whether the resolution lies with a need for higher orders, a need to refit correlation functions, large power-law corrections in the region of moderate $Q$~\cite{Liu:2019srj}, 
or still other factors that are not yet understood remains unclear.   

An important early step toward clarifying the issues is an examination of trends in standard methods of calculation in the large transverse momentum region. Motivated by 
this, we have examined the simplest LO calculation 
relevant for large deviation from the back-to-back region in detail. 
Agreement with Monte Carlo-generated distributions at large $Q$ supports the general validity of such calculations. 
However, when comparing the result in the intermediate transverse momentum region with expectations obtained from TMD fragmentation functions, we find trends reminiscent of those discussed above for SIDIS and Drell-Yan scattering at lower $Q$. Namely, the collinear factorization calculation appears to be overly suppressed. 
We view this as significant motivation to study the intermediate transverse momentum region both experimentally and theoretically. An advantage in the 
$e^+ e^-$ annihilation is the larger value of $Q$ relative to processes like semi-inclusive deep inelastic scattering.

While we have focused on the large transverse momentum limit, the  observations above are relevant to other kinematical 
regions such as small transverse momentum, as well as to polarization dependent observables, and their physical interpretation, since the 
detailed shape of the transverse momentum distributions for any region depend on the 
transitions to other regions. 

It is important to note that order 
$\alpha_s^2$ corrections can be quite 
large~\cite{Daleo:2004pn,Kniehl:2004hf,Gonzalez-Hernandez:2018ipj,Wang:2019bvb}, and we plan to 
address these in future studies, though generally higher order effects have not been sufficient in other processes to 
eliminate tension. 
Keeping this in mind, it is worthwhile nevertheless to speculate on other possible resolutions. 
One is that the hard scale $Q$ might be too low for a 
simplistic division of transverse momentum into regions such as discussed in \sref{qthardness}. It is true that as $Q$ gets smaller, the separation 
between large and small transverse momentum becomes squeezed, and it is possible that the standard 
methods for treating the transition between separately well defined regions is inapplicable. As a 
hard scale, however, $Q \sim 12$~GeV is well above energies that are normally understood to be near to the 
lower limits of applicability of standard perturbation theory methods (typical scales for SIDIS measurements are around 
$Q \sim 2$~GeV, for example). Another possibility is that fragmentation functions in the large $\zeta$ 
range probed at large $\Tsc{q}{}$ are not sufficiently constrained.  An important next step is to 
determine  whether the description of large transverse momentum processes generally can be 
improved via a simultaneous analysis of multiple processes at moderate $Q$ with simple and well-
established collinear factorization treatments. We plan to investigate this in future work.

%%%%%%%%%%%%%%%%%%%%%%%%%%%%%%%%%%%%%%%%%%%%%%%%%%%
\appendix

%%%%%%%%%%%%%%%%%%%%%%%%%%%%%%%%%%%%%%%%%%%%%%%%%%%
%\newpage

\section{Variables Changes}
\label{a.cov}

The left hand side of \eref{tensors} can be rewritten as
\begin{equation}
\label{e.cross1}
\frac{\left|\3{p}_A\right|}{\left|\3{p}_B\right|}\frac{\diff{\sigma_{AB}}}{\diff{\left|\3{p}_B\right|}\diff{\Omega_B}\diff^3\3{p}_A}
\end{equation}
Change of variables is easiest in a center of mass frame where $p_B$
is on the z-axis.  In this frame, the hadron momenta in terms of $Q$,
$\Tsc{q}{}$, $z _A$, and $z_B$ (in Cartesian coordinates) are:
\begin{align}
p_A&=\parz{\frac{z
_A}{2Q}\parz{Q^2+\Tscsq{q}{}},-z_A\T{q},-\frac{z_A}{2Q}\parz{Q^2-\Tscsq{q}{}}} \\
p_B&=\parz{\frac{z
_B}{2Q}\parz{Q^2+\Tscsq{q}{}},\T{0}{},\frac{z_B}{2Q}\parz{Q^2+\Tscsq{q}{}}}.
\end{align}
and the lepton momentum $l$ is:
\begin{equation}
l=\parz{\frac{Q}{2},\frac{Q}{2}\sin\theta\cos\phi,\frac{Q}{2}\sin\theta\sin\phi,\frac{Q}{2}\cos\theta}.
\end{equation}
Therefore:
\begin{align}
\left|\3{p}_A\right|&=\frac{z_A}{2Q}\parz{Q^2+\Tscsq{q}{}} \no
\left|\3{p}_B\right|&=\frac{z_B}{2Q}\parz{Q^2+\Tscsq{q}{}} \no
\diff^3\3{p}_A\diff{\left|\3{p}_B\right|}&=\frac{\Tsc{q}{}(Q^2+\Tsc{q}{}^2)^2z_A^2}{4Q^2}\diff{z_A}\diff{z_B}\diff\Tsc{q}{}\diff{\phi_A}\no
\diff{\Omega_B}&=\diff{\cos\theta}\diff{\phi}
\end{align}
After integrating over $\phi_A$, \eref{tensors} then becomes:
\begin{equation}
\frac{\diff{\sigma_{AB}}}{\diff{z_A}\diff{z_B}\diff{\Tsc{q}{}}\diff{\cos\theta}\diff{\phi}}=\frac{\alpha_{{\rm em}}^2z_Az_B\parz{Q^2+\Tsc{q}{}^2}^2\Tsc{q}{}}{16\pi^2{Q}^8}L_{\mu \nu} W^{\mu \nu}.
\end{equation}

\section{Fixed Order Expressions}
\label{a.fo}

The partonic structure functions $\widehat{W}_{T,ij}$ and $\widehat{W}_{L,ij}$ can be obtained by contracting the extraction tensors
(\eref{extens}) with the partonic tensor $\widehat{W}^{\mu\nu}$. The relation between the 
partonic tensor and the squared amplitude of the hard
part is:
\begin{equation}
\widehat{W}^{\mu\nu}_{ij}=4 \pi^3 \int \frac{\diff{^3\3{k}}_C}{2 k_C^0 (2 \pi)^3}  
\delta^{(4)}\parz{q - k_A - k_B - k_C}|\mathcal{\widehat{M}}|^{2,\mu\nu}_{ij}  = \frac{1}{2} \delta_+(k_C^2)|\mathcal{\widehat{M}}|^{2,\mu\nu}_{ij} \, .
\end{equation}
The resulting partonic cross sections are 
\begin{subequations}
\label{e.fopart}
\begin{align}
%-------------------------------------
\frac{\diff{\hat{\sigma}}_{q\bar{q}}}{\diff{\hat{z}}_A\diff{\hat{z}}_B\diff{\Tsc{q}{}}}
=\frac{\diff{\hat{\sigma}}_{\bar{q}q}}{\diff{\hat{z}}_A\diff{\hat{z}}_B\diff{\Tsc{q}{}}}
=&\frac{8 \alpha_{\rm em}^2 \alpha_s e_q^2 \hat{z}_A \hat{z}_B \delta\parz{k_C^2}\Tsc{q}{} \parz{Q^2+\Tscsq{q}{}}^3 \parz{6 Q^2+5 \Tscsq{q}{}} \parz{\hat{z}_A^2+\hat{z}_B^2}}{9 Q^6 \parz{Q^2 (\hat{z}_A-1)+\Tscsq{q}{} \hat{z}_A} \parz{Q^2 (\hat{z}_B-1)+\Tscsq{q}{} \hat{z}_B}}\\
%-------------------------------------
\frac{\diff{\hat{\sigma}}_{qg}}{\diff{\hat{z}}_A\diff{\hat{z}}_B\diff{\Tsc{q}{}}}
=\frac{\diff{\hat{\sigma}}_{\bar{q}g}}{\diff{\hat{z}}_A\diff{\hat{z}}_B\diff{\Tsc{q}{}}}
=&-8 \alpha_{\rm em}^2 \alpha_s e_q^2 \hat{z}_A \hat{z}_B\delta\parz{k_C^2} \Tsc{q}{}\parz{Q^2+\Tscsq{q}{}}^2\left[2 Q^4 \parz{14+3\hat{z}_B^2-14\hat{z}_B+2\hat{z}_A\parz{3\hat{z}_A+4\hat{z}_B-7}}\right.\no
&\left.+5\Tsc{q}{}^4\parz{\hat{z}_B^2+2\hat{z}_B\hat{z}_A+2\hat{z}_A^2}+Q^2\Tscsq{q}{}\parz{11\hat{z}_B^2-28\hat{z}_B+2\hat{z}_A\parz{11\hat{z}_A+13\hat{z}_B-14}}\right]\no
&\Bigg/%\div
\parz{9 Q^6 \parz{Q^2 ({\hat{z}}_A-1)+\Tscsq{q}{} {\hat{z}}_A} \parz{Q^2 (\hat{z}_B+{\hat{z}}_A-1)+\Tscsq{q}{} (\hat{z}_B+{\hat{z}}_A)}}\\
%-------------------------------------
\frac{\diff{\hat{\sigma}}_{gq}}{\diff{\hat{z}}_A\diff{\hat{z}}_B\diff{\Tsc{q}{}}}
=\frac{\diff{\hat{\sigma}}_{g\bar{q}}}{\diff{\hat{z}}_A\diff{\hat{z}}_B\diff{\Tsc{q}{}}}
=&-8 \alpha_{\rm em}^2 \alpha_s e_q^2 \hat{z}_A \hat{z}_B \delta\parz{k_C^2}\Tsc{q}{}\parz{Q^2+\Tscsq{q}{}}^2 \left[2 Q^4 \parz{14+3\hat{z}_A^2-14\hat{z}_A+2\hat{z}_B\parz{3\hat{z}_B+4\hat{z}_A-7}}\right.\no
&\left.+5\Tsc{q}{}^4\parz{\hat{z}_A^2+2\hat{z}_A\hat{z}_B+2\hat{z}_B^2}+Q^2\Tscsq{q}{}\parz{11\hat{z}_A^2-28\hat{z}_A+2\hat{z}_B\parz{11\hat{z}_B+13\hat{z}_A-14}}\right]\no
& \Bigg/
\parz{9 Q^6 \parz{Q^2 ({\hat{z}}_B-1)+\Tscsq{q}{} {\hat{z}}_B} \parz{Q^2 (\hat{z}_A+{\hat{z}}_B-1)+\Tscsq{q}{} (\hat{z}_A+{\hat{z}}_B)}}
%-------------------------------------
\end{align}
\end{subequations}

%%%%%%%%%%%%%%%%%%%%%%%%%%%%%%%%%%%%%%%%%%%%%%%%%%%

\begin{acknowledgments}
Discussions with Elena Boglione, Markus Diefenthaler, J.~Osvaldo~Gonzalez-Hernandez, Charlotte van Hulse, Ralf Seidl and Anselm Vossen are gratefully
acknowledged.  
T.~Rogers, E. Moffat, and N.~Sato were supported by the U.S. Department of Energy, Office of 
Science, Office of Nuclear Physics, under Award Number DE-SC0018106.
A.~Signori  acknowledges  support  from  the U.S. Department of
Energy, Office of Science, Office of Nuclear Physics, contract no.
DE-AC02-06CH11357.
This work 
was also supported by the DOE Contract No. DE- AC05-06OR23177, under which 
Jefferson Science Associates, LLC operates Jefferson Lab. 
\end{acknowledgments}

%%%%%%%%%%%%%%%%%%%%%%%%%%%%%%%%%%%%%%%%%%%%%%%%%%%
\bibliography{bibliography}

\end{document}